\RequirePackage{lineno}
\documentclass[aps,prc,twocolumn, floatfix]{revtex4}
\usepackage{graphicx}
\usepackage{dcolumn}
\usepackage{bm}

\begin{document}


\title{Isospin blocking and its effects in heavy-ion collisions}

\author{Ya-Fei Guo$^{1,2,3,4}$}
\author{Gao-Chan Yong$^{3,4}$}
\email[]{yonggaochan@impcas.ac.cn}

\affiliation{$^1$Institute of Particle and Nuclear Physics, Henan Normal University, Xinxiang 453007, China\\
$^2$School of Physics, Henan Normal University, Xinxiang 453007, China\\
$^3$Institute of Modern Physics, Chinese Academy of Sciences, Lanzhou 730000, China\\
$^4$School of Nuclear Science and Technology, University of Chinese Academy of Sciences, Beijing 100049, China}

\begin{abstract}

A concept of \emph{isospin blocking} in the process of isospin diffusion in heavy-ion collisions is  raised. Generally, in the process of isospin diffusion, isospin asymmetry would diffuse from the place with large asymmetry to the place with small asymmetry. However, our study shows that the isospin diffusion could be blocked in case the local value of the symmetry energy is larger. We dub this phenomenon ``isospin blocking''. To check this behavior, in the framework of an Isospin-dependent Boltzmann-Uehling-Uhlenbeck (IBUU) transport model, isospin diffusions in the isotope Sn+Sn reactions at 270 MeV/nucleon are studied. It is shown that the value of the after-diffusion asymmetry is distinctly blocked if the local symmetry energy is large. The effects of the isospin blocking on the isospin asymmetry of dilute and dense matter and the final $\pi^-/\pi^+$ ratio in heavy ion collisions are demonstrated.

\end{abstract}

\maketitle


The properties of general nuclear matter can be described by the equation of state (EoS) with density $\rho$ and isospin asymmetry $\delta=(\rho_n-\rho_p)/\rho$ as \cite{bardo}
\begin{equation}\label{equ0}
E(\rho,\delta)=E(\rho, 0)+E_{sym}(\rho)\delta^2+O(\delta^4),
\end{equation}
where $E_{sym}(\rho)$ is the density-dependent nuclear symmetry energy, which is a measure of the energy cost to make nuclear systems more neutron rich. The symmetry energy is known to play crucial role in studies of both nuclear physics \cite{1Da,2Ba,3Li,4Tr} and astrophysics \cite{5Lat,6Ab,7Ts}. Therefore the density dependence of the nuclear symmetry energy has been extensively studied both experimentally and theoretically \cite{8vid,9Ts,10Mo,11Do,12Da,13Mo,14Ad,fopi16}. The value of the symmetry energy and its slope around saturation density have nowadays been roughly constrained \cite{15Li,yong2021,24Es,chenlw2021,reed2021} while its high-density behavior is still controversial. The high-density symmetry energy is of great importance in the studies of binary neutron star mergers \cite{16GWth,17GW170817} and supernova explosions \cite{super}.

One of the main methods to constrain the high-density symmetry energy is through heavy ion collisions in terrestrial laboratories \cite{2Ba,3Li}. In the collisions, the colliding target and projectile nuclei are compressed beyond saturation density, meanwhile $\pi$ mesons could emit from the dense matter. The emitting pionic particles from the dense matter carry information on the high-density symmetry energy \cite{18Ik}. Because the $\pi^-/\pi^+$ ratio strongly depends on isospin asymmetry of the participant region of heavy-ion collisions \cite{19Li1,20Ho,21Ik1}, it is frequently studied as a potential observable of the high-density symmetry energy.

To check if pions produced in heavy-ion collisions really carry the information on the high-density symmetry energy, the characteristic density of pion and the weighted density of $\Delta$-force are studied recently \cite{22Liu}. Besides such \emph{indirect} studying methods, the more direct methods on whether the produced pions in heavy-ion collisions probe the high-density symmetry energy were in fact carried out previously by switching on or off the symmetry energy in specific density regions \cite{34yong20191,32Liu1} and more recently a more advanced tracer technique was carried out \cite{23Gao}.

The main purpose of the recent experiments of the isotope Sn+Sn reactions at 270 MeV/nucleon conducted by the S$\pi$RIT collaboration is to probe the high-density symmetry energy by the $\pi^-/\pi^+$ ratio \cite{24Es,25Jh}. So it is urgently necessary to investigate if the $\pi^-/\pi^+$ ratio in such experiments really probes the high-density symmetry energy \cite{cozmat,wolter22}.

Physically, when a nuclear system is brought out of isospin equilibrium, the gradient of the symmetry energy could result in a net flow of isospin asymmetry, thus isospin diffusion occurs \cite{diff0,shi2000,shi2003,diff1,diff2}. Keeping the total asymmetry of the nuclear system fixed, if a dilute phase is in isospin equilibrium with a dense phase, to minimize total free energy, the equilibrium condition
\begin{equation}\label{equ1}
E_{sym}^{L}\delta^{L} = E_{sym}^{G}\delta^{G}
\end{equation}
should be satisfied. Where $E_{sym}^{L}, E_{sym}^{G}, \delta^{L}, \delta^{G}$ are the symmetry energies and asymmetries of liquid and gas phases, respectively \cite{shi2000,linpa2002}. From Eq.~(\ref{equ1}), one can deduce
\begin{equation}\label{equ2}
\delta^{L,G} = \frac{E_{sym}^{G,L}}{E_{sym}^{L,G}}\delta^{G,L},
\end{equation}
thus the asymmetry of dense (or dilute) phase is affected by the symmetry energies from both dense and dilute phases. Based on an Isospin-dependent transport model, in this Letter, we show how the high-density isospin asymmetry is affected by the low-density symmetry energy. More over, such effects on the $\pi^-/\pi^+$ ratio in the isotope Sn+Sn reactions at 270 MeV/nucleon are demonstrated accordingly.


In the present studies, we use an Isospin-dependent Boltzmann-Uehling-Uhlenbeck (IBUU) transport model with the Skyrme-type parametrization for the isoscalar term of the mean field potential, i.e., $U(\rho)$ = -232$(\rho/\rho_0)$ + 179$(\rho/\rho_0)^{1.3}$ \cite{34yong20191}. For the isovector part of the mean field, a form corresponding to the density-dependent symmetry energy $E_{sym}(\rho)$ = 32$(\rho/\rho_0)^{\gamma}$ is used \cite{34yong20191}. For the convenience of the present studies, at high densities ($\rho \geq \rho_0$), a mildly soft symmetry energy form with $\gamma$ = 0.5 is fixed. To study the isospin blocking phenomenon, we vary the symmetry energy at low densities ($\rho < \rho_0$), i.e., the forms with $\gamma$ = 0.5 (corresponding to large value of the symmetry energy) and $\gamma$ = 1 (corresponding to small value of the symmetry energy) are used, respectively. Note here that our presentation is mainly qualitative, since a large variety of the symmetry energy at low or high densities used in this study can demonstrate the core of the problem easily. Therefore, we in the following use a momentum-independent form of the single nucleon potential.

\begin{figure}[tbh]
\begin{center}
\includegraphics[width=0.45\textwidth]{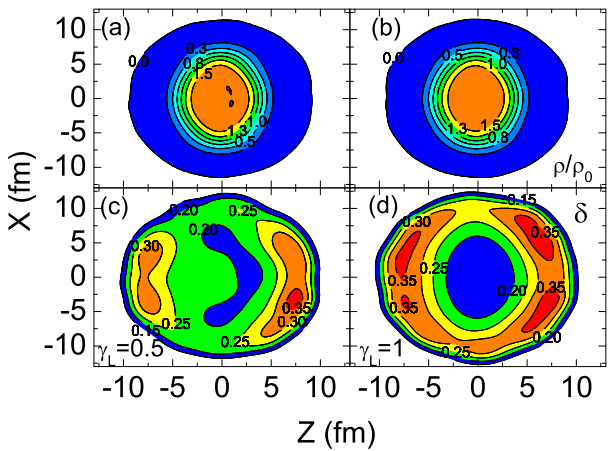}
\end{center}
\caption{Spatial distributions of density (upper panels) and isospin asymmetry (lower panels) at maximum compression stage in the central $^{132}$Sn+$^{124}$Sn reactions at incident beam energy 270 MeV/nucleon. The left (right) panel corresponds to large (small) symmetry energy at low densities ($\rho < \rho_0$).
}\label{cont}
\end{figure}
Figure~\ref{cont} shows spatial distributions of density and isospin asymmetry at maximum compression stage in the central $^{132}$Sn+$^{124}$Sn reactions at incident beam energy 270 MeV/nucleon. It is seen that isospin diffusion has already occurred at the maximum compression stage and larger asymmetries locate at the low-density regions. This is understandable from the chemical equilibrium condition during the isospin fractionation \cite{shi2000,shi2003,linpa2002}. More importantly, comparing panel (c) to panel (d), it is clearly shown that at lower densities the weak (strong) symmetry energy causes large (small) isospin asymmetry. This indicates the occurrence of isospin blocking at lower densities if the symmetry energy is strong. From Figure~\ref{cont}, one can also see that in the central collisions larger isospin asymmetry of nucleon gas roughly locates in the beam direction (positive $z$ direction).

\begin{figure}[tbh]
\begin{center}
\includegraphics[width=0.45\textwidth]{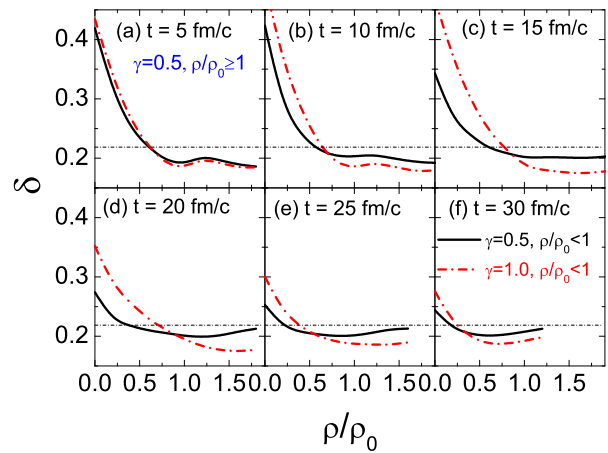}
\end{center}
\caption{Time evolution of the isospin asymmetry as a function of density in the central $^{132}$Sn+$^{124}$Sn reactions at 270 MeV/nucleon with large ($\gamma$ = 0.5) and small ($\gamma$ = 1) symmetry energies at low densities. The dash-dot line stands for the isospin asymmetry of the system.}\label{whev}
\end{figure}
Figure~\ref{whev} shows the evolution of the isospin asymmetry at different densities in the $^{132}$Sn+$^{124}$Sn reactions at 270 MeV/nucleon with different symmetry energies at low densities. The most prominent characteristic of the distribution of isospin asymmetry as a function of density is that the isospin asymmetry is generally larger (smaller) at low (high) densities \cite{shi2000,linpa2002}. At the beginning stage of the reaction (t = 5 fm/c), larger isospin asymmetry at low densities is mainly due to the effects of neutron skin. As compression goes on, isospin diffusion occurs. The high-density symmetry energy tends to repel neutrons to the low-density region whereas the larger symmetry energy at low-density region blocks neutrons going into the low-density region. One sees the effects of isospin blocking become visible in the compression stage of the reaction from t = 10 fm/c to t = 20 fm/c. This is also the stage in which the symmetry energy shows maximum effects on the reaction dynamics. This is the reason why the pre-equilibrium nucleon emission is frequently used to probe the high-density symmetry energy \cite{2Ba,3Li}. At the expansion stage, the dense matter dissociates thus the isospin asymmetry at low densities tends to the system's asymmetry. Therefore the isospin blocking gradually disappears at the expansion stage in heavy-ion collisions. From Figure~\ref{whev}, one can also see that the isospin blocking not only affects the isospin asymmetry of dilute nuclear matter, i.e., nucleon gas, but also affects the isospin asymmetry of dense matter formed in heavy-ion collisions.

\begin{figure}[tbh]
\begin{center}
\includegraphics[width=0.45\textwidth]{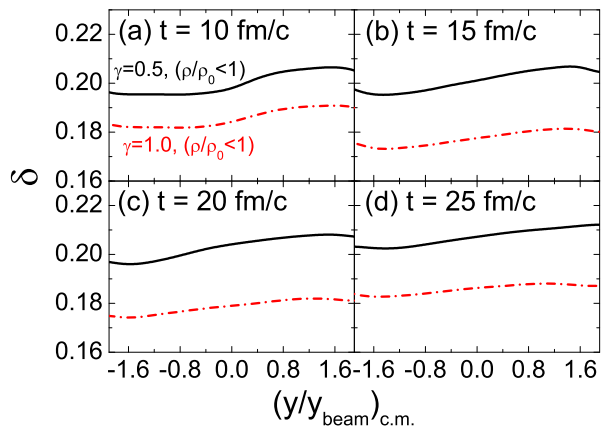}
\end{center}
\caption{Same as Figure~\ref{whev}, but for the isospin asymmetry as a function of normalized
center-of-mass rapidity in the high-density region ($\rho \geq \rho_0$). Fixing the symmetry energy at high densities, the  symmetry energy varies from weak ($\gamma$ = 1) to strong ($\gamma$ = 0.5) at low densities.}\label{deny}
\end{figure}
Generally, there are two mechanisms to probe the symmetry energy, one is probing the isospin asymmetry of dilute nuclear matter, i.e., nucleon gas. This method has been extensively utilized to probe the density-dependent symmetry energy, such as the frequently used free neutron to proton n/p ratio and isospin flow in heavy-ion collisions \cite{guo2019,ditoro2005,ditoro2010}. This is a direct observable since the symmetry potential directly acts on the emitting nucleons in the reaction dynamics. The other method is probing the isospin asymmetry of dense matter, such as the intermediate mass fragments (IMF) \cite{ditoro2008,3Li,imf1,imf2,imf3}, the $\pi^-/\pi^+$ ratio \cite{19Li1,20Ho,21Ik1,liba2005,24Es,25Jh}, the $K^0/K^+$ ratio \cite{kaon2006,rkaon07} in heavy-ion collisions. This method measures the isospin asymmetry of dense matter through colliding effects of nucleons in dense matter, thus indirectly reflects the isospin asymmetry of dense matter.

Figure~\ref{deny} shows the isospin asymmetry of dense matter as a function of rapidity. It is clearly seen that the isospin asymmetry of dense matter is affected by the symmetry energy at low densities. The strong low-density symmetry energy increases the isospin asymmetry of dense matter while the weak low-density symmetry energy decreases the isospin asymmetry of dense matter. This is so called the recoil effect of the isospin blocking phenomenon at low densities. The recoil effects are shown to reach a maximum at maximum compression (t = 15, 20 fm/c) in the collisions. Such behavior is consistent with that shown in Figure~\ref{whev}, where the isospin blockings are clearly shown at low densities. By comparing the negative with the positive rapidities, it is seen that the isospin transparency is less affected by the isospin blockings. The larger isospin asymmetry shown in the positive rapidity region is due to the larger asymmetry of the projectile $^{132}$Sn than the target $^{124}$Sn.

\begin{figure}[t!]
\begin{center}
\includegraphics[width=0.42\textwidth]{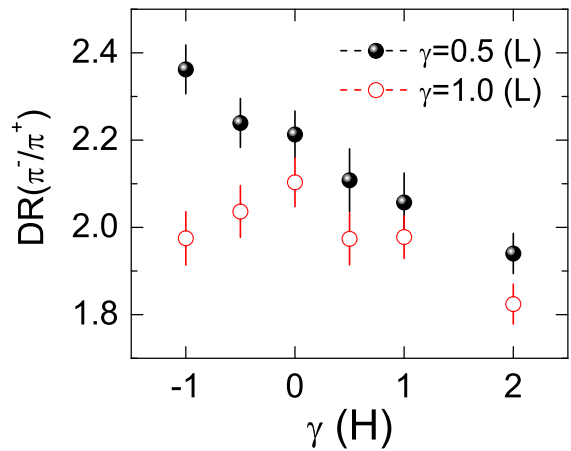}
\end{center}
\caption{Double $\pi^-/\pi^+$ ratio as a function of high-density symmetry energy parameter $\gamma$ with weak ($\gamma$ = 1) and strong ($\gamma$ = 0.5) symmetry energies at low densities in the $^{132}$Sn+$^{124}$Sn and $^{108}$Sn+$^{112}$Sn reactions at 270 MeV/nucleon.}\label{drpi}
\end{figure}
To determine the high-density symmetry energy, the single and double $\pi^-/\pi^+$ ratios in heavy-ion collisions are frequently used as potential observables. Very recently, both the single and double $\pi^-/\pi^+$ ratios in $^{132}$Sn+$^{124}$Sn and $^{108}$Sn+$^{112}$Sn reactions at 270 MeV/nucleon were measured by the S$\pi$RIT collaboration. The $\pi^-/\pi^+$ ratio in heavy-ion collisions is an indirect reflection of the isospin asymmetry of compressed dense matter and the isospin asymmetry of dense matter is evidently affected by the recoil effects of the isospin blocking as shown in Figure~\ref{deny}, but to what extend does the isospin blocking affect the $\pi^-/\pi^+$ ratio in heavy-ion collisions? To this end, the double $\pi^-/\pi^+$ ratio in the $^{132}$Sn+$^{124}$Sn and $^{108}$Sn+$^{112}$Sn reactions at 270 MeV/nucleon are studied with different high-density symmetry energies. The double $\pi^-/\pi^+$ ratio reads
\begin{equation}
DR(\pi^-/\pi^+)=\frac{(\pi^-/\pi^+)_{^{132}Sn+^{124}Sn}}{(\pi^-/\pi^+)_{^{108}Sn+^{112}Sn}}.
\end{equation}
From Figure~\ref{drpi}, it is expectedly seen that as the high-density symmetry energy becomes soft, the value of the double $\pi^-/\pi^+$ ratio increases monotonously with whether a weak or a strong symmetry energy at low densities. But this monotonous upward behavior turns to decrease when a flat high-density symmetry energy and a weak symmetry energy at low densities are simultaneously used. Therefore, the double $\pi^-/\pi^+$ ratio in the $^{132}$Sn+$^{124}$Sn and $^{108}$Sn+$^{112}$Sn reactions at 270 MeV/nucleon mainly probes the high-density symmetry energy except the high-density symmetry energy does not increase with the density \cite{blind2018}. In Ref.~\cite{blind2018}, it is shown that if the high-density symmetry energy is less density-dependent, then the observable does not reflect the information of the high-density symmetry energy (i.e., the high-density symmetry energy's effects could be zero!), in this case, the low-density symmetry energy has chance to affect the observable much than the high-density symmetry energy. The recoil effects of the isospin blocking (from the low-density region) do affect the value of the double $\pi^-/\pi^+$ ratio in heavy-ion collisions especially a rather soft high-density symmetry energy is supposed.

Isospin blocking reveals the physical mechanism that how the variety of low-density symmetry energy plays a role in probing the high-density symmetry energy by using some sensitive observables. In principle, isospin blocking should exist extensively in isospin diffusion or fractionation in liquid-gas phase transition only if the density region is broad, not just limited to the existence of super-dense matter. The studies relevant to isospin diffusion in the literature usually assumes a definite density-dependent symmetry energy from low to high densities, thus there is no need to raise the isospin blocking concept. The isospin blocking may need to be checked if one wants to probe the symmetry energy at relatively higher densities (does not have to be greater than normal nuclear density) in case the low-density symmetry energy has room to change.


In conclusions, a concept of isospin blocking in the process of isospin diffusion in heavy-ion collisions is raised while probing the high-density symmetry energy in heavy-ion collisions. The effects of so called isospin blocking on the isospin asymmetries of both dilute matter and dense matter are demonstrated. As a consequence, the double $\pi^-/\pi^+$ ratio in the Sn+Sn reactions recently measured by S$\pi$RIT collaboration is not only a reflection of the high-density symmetry energy, but also a reflection of the recoil effects of the isospin blocking at low densities. The present study may help one to understand the physical mechanism of interplay of low and high-density symmetry energies in the process of isospin diffusion and to constrain the high-density symmetry energy in heavy-ion collisions by using symmetry-energy-sensitive observables.


This work is supported in part by the Strategic Priority Research Program of Chinese Academy of Sciences, Grant No. XDB34030000 and the National Natural Science Foundation of China under Grant No. 11775275 and the Program for Innovative Research Team (in Science and Technology) in University of Henan Province, China (No. 21IRTSTHN011).

\end{document}